\documentclass[%preprint,
prd,aps,twocolumn, 
showpacs,tightenlines,nofootinbib,preprintnumbers, superscriptaddress
]{revtex4}
\usepackage{graphicx}
\usepackage{amssymb,amsmath,latexsym}
\usepackage{amsfonts}
\usepackage{cancel}
\usepackage{color}
\input{colordvi.tex}
\input{epsf}
\usepackage{epsf}
\usepackage{graphicx,epsfig}
\usepackage{bm}
\usepackage{latexsym,float}

\newcommand{\beqa}{\begin{eqnarray}}
\newcommand{\eeqa}{\end{eqnarray}}

\newcommand{\solar}{\odot}
\newcommand{\beq}{\begin{equation}}  \newcommand{\eeq}{\end{equation}}
\newcommand{\bef}{\begin{figure}}  \newcommand{\eef}{\end{figure}}
\newcommand{\bec}{\begin{center}}  \newcommand{\eec}{\end{center}}

\newcommand {\ga} {\ {\raise-.5ex\hbox{$\buildrel>\over\sim$}}\ }
\newcommand {\la} {\ {\raise-.5ex\hbox{$\buildrel<\over\sim$}}\ }

\begin{document}
%\draft

\title{Shadows of Colliding Black Holes}

\author{Daisuke Nitta}
\affiliation{Department of Physics, Nagoya
  University, Chikusa, Nagoya 464-8602, Japan} 
\author{Takeshi Chiba}
\affiliation{Department of Physics, \\
College of Humanities and Sciences, \\
Nihon University, \\
Tokyo 156-8550, Japan}
\author{Naoshi Sugiyama} 
\affiliation{Department of Physics, Nagoya
  University, Chikusa, Nagoya 464-8602, Japan} \affiliation{Institute
  for Physics and Mathematics of the Universe, University of Tokyo,
  Chiba 277-8582, Japan}

\date{\today}


\begin{abstract}
We compute the shadows of colliding black holes using the Kastor-Traschen  
cosmological multiblack hole solution that is  
an exact solution describing the collision of maximally charged black holes 
with a positive cosmological constant. 
We find that in addition to the shadow of each black hole, an eyebrowlike structure appears  as 
the black holes come close to each other.
These features can be used as probes to find 
the multiblack hole system at the final stage of its merger process.
%The Kastor-Traschen solution is a multi black hole solution with non-zero cosmological constant. 
%In this solution, the black holes move with \red{cosmic} expansion or contraction 
%so that a merger of black hole is naturally described. 
%Therefore the shadows using this solution can be useful tool for investigating the black hole collision. 
%We find that the shadows have some features which correspond to the black hole 
%mergers such as eyelash-like structures.
%We show that obtained results can give observational suggestions to the realistic black hole %merger in the future.

\end{abstract}

\maketitle

\section{Introduction}

The observational evidence for the existence of black holes is
mounting (see \cite{narayan} for recent reviews).  There are many
stellar mass black holes found in the Galaxy, while one of them, Sgr
${\rm A}^*$, turns out to be a super massive black hole (SMBH) with
$4.3\times 10^6$ solar mass~\cite{ghez,gillessen}.  
It turns out most galaxies and active galactic nuclei have
at least one SMBH whose mass shows strong correlation with the mass of
the spheroid component of the galaxy~\cite{kormendy,magorrian,merritt}.  
The hierarchical clustering
scenario suggests such a spheroid component is formed due to a merger of
smaller galaxies. Therefore it may be natural to consider the
formation of a SMBH is also taking place by the merger process of smaller
black holes.

Recently, the evidence for the existence of a binary black holes is  
provided by observing the orbital motions of stars  
in a galaxy by radio interferometers \cite{sudou}. 
Moreover, from the detection of a signal periodicity in light curves, 
it is claimed that the binary black holes will coalesce within 500 yr \cite{iguchi}.  
However, the direct evidence of black holes is still lacking. 
We need unambiguous proof that this object is indeed a binary black holes. 
Since a black hole is defined as an object with the event horizon, we should 
search for phenomena associated with the existence of the event horizon. 

To a distant observer, the event horizons cast shadows due to the bending of light 
by the black holes \cite{falcke}.  Observing these shadows should be  compelling 
evidence of a coalescing black holes. 

As a first step toward the study of a realistic black hole binary, we calculate 
the shadow of the Kastor-Traschen cosmological multiblack hole solutions \cite{kt}. 
The Kastor-Traschen solution is an exact solution describing the merger of maximally charged black holes with a positive cosmological constant.  
Although admittedly the solution is unrealistic, it is an exact and analytic solution and 
hence allows us to study numerically photon orbits accurately. 
We expect some of the features of the shadows would persist for a 
more realistic black hole binary 
since, at least for a single black hole, the charge of black holes has little effect on the apparent shape of the shadow \cite{takahasi}. 

\section{Kastor-Traschen solution}

In this section, we will briefly review the Kastor-Traschen(KT) solution.
The KT solution is a dynamic multiblack hole solution 
%(some other solutions have been suggested, see \cite{mp,gibbons}) 
parameterized by $n$ masses $m_i$  
%, locations of the black holes $(x_i,y_i,z_i)$
and the positive cosmological constant $\Lambda$ (see also \cite{gibbons}). 
Each black hole has charge $Q_i$ equals to its mass $m_i$ 
(we use the geometrical units, $G=c=1$). 
In case of a single black hole, this solution corresponds to the Reissner-Nordstrom-de Sitter solution
with charge equals to its mass. It can be reduced to the Majumdar-Papapetrou solution when $\Lambda=0$ \cite{mp}.

The metric in the cosmological coordinate is given by
\begin{eqnarray}
&&ds^2=-a^2\Omega^{-2}d{\tau}^2+a^2\Omega^2(dx^2+dy^2+dz^2),
\nonumber\\
&& a=e^{Ht}=-\frac{1}{H\tau},\quad H=\pm \sqrt{\frac{\Lambda}{3}}\quad\Omega=1+\sum_{i}\frac{m_i}{ar_i},\nonumber\\ 
&&r_i\equiv \sqrt{(x-x_i)^2+(y-y_i)^2+(z-z_i)^2},
\label{eq:multi}
\end{eqnarray}
where, $\tau$ and $t$ denote conformal time and physical time respectively.
Here, $H>0$ ($H<0$) corresponds to expansion (contraction). 

In the Majumdar-Papapetrou solution, the black holes can stay at the rest frame as if their gravity balance with electrostatic repulsions. 
Similarly, in the KT solution, the gravity of the black holes balance with their electrostatic repulsions,  
%due to their charge which equals to their masses.
while the black holes comove with cosmic expansion.
%the expansion or contraction only.

The solution has some interesting features. Let us consider two
extreme situations.  First, we imagine the case that black holes are
far enough from each other.  If distances of black holes are at least
larger than $1/|H|$, then each black hole can be treated as a single black
hole.  Second, we consider the case that all black holes are close
enough to each other.  If all black holes are located within the black
hole horizon of total mass $\sum_{i}m_i$, this system can be
considered as a single black hole~\cite{nakao}.  Accordingly, one can
see that the KT solution describes the black hole collision because in
the contracting coordinate, it starts from a group of single black
holes and ends up with a single black hole.
%(or the reverse process in the expanding coordinate).
% This fact implies that the KT solution contains process which of the event horizons \cite{nakao}, this solution describes black hole collisions.

\section{Shadow of a single black hole}

We begin by computing the shadow of a single black hole to understand
some asymptotic behaviors.  The KT solution in case of a single black
hole is equivalent to the extreme Reissner-Nordstrom-de Sitter solution. This solution can be
rewritten in the static coordinate. The metric is then given by
\begin{eqnarray}
&&ds^2=-VdT^2+V^{-1}dR^2+R^2(d\theta^2+\sin^2\theta d\phi^2)
,\label{eq:staticRN} \nonumber\\
&&V(R)=\left(1-\frac{M}{R}\right)^2-H^2R^2.
\end{eqnarray}
Taking $V=0$, we obtain event horizons which are given by
\begin{eqnarray}
R_{\pm}=\frac{1}{2|H|}(1\pm \sqrt{1-4M|H|}), \label{eq:statichorizons}
\end{eqnarray}
where $R_+$ and $R_{-}$ denote cosmological and black hole horizons,
respectively.   The cosmological horizon has a similar feature of the
horizon in de Sitter space-time since it becomes a future (past)
horizon if the Universe is contracting (expanding).  On the other hand, 
the black hole horizon is a usual black hole horizon in the RDdS solution.

We define a momentum of a photon using the affine parameter $\lambda$ as $P^{\mu}\equiv dx^{\mu}/d\lambda$.
%where $\mu=T,R,\theta,\phi$.
%The null condition is given by
%\begin{eqnarray}
%g_{\mu\nu}P^{\mu}P^{\nu}=0,\quad P^{\mu}\equiv \frac{d x^{\mu}}{d\lambda}.
%\end{eqnarray}
Here, $P_T$ and $P_{\phi}$ are constants corresponding to the time shift and the rotational symmetry.
Using the null condition, we obtain a geodesic equation at $\theta =\pi/2$
\begin{eqnarray}
{\left( \frac{dR}{d\lambda}  \right)}^2+VR^{-2}b^2=1,\label{eq:enecon}
\end{eqnarray}
where, $b\equiv P_{\phi}/P_{T}$ denotes the impact parameter.  The
``effective potential" $b^2V/R^2$ has a local maximum at $R=2M$. A
sphere with this radius of $R=2M$ is known as the ``photon sphere"
inside of which the photon orbits become unstable.
%where photons wind around the black hole.  
Note that the orbits of these photons are unstable.  
%which is the region
% where the photons comes from the black hole can never escape the black hole due to its gravity. 
One can hence find that the critical value of the impact parameter $b_c$ is given by
\begin{eqnarray}
b_c=\frac{4M}{\sqrt{1-16M^2H^2}}.\label{eq:impact}
\end{eqnarray}
If $b\le b_c$, the photons are captured by the black hole.
%On the boundary of $b_c$, there are three kinds of photon orbits; 
%A scattering orbit for $b>b_c$, in which the photons can travel from light sources to the observer, 
%A captured orbit for $b<b_c$, in which the photons always fall into the black hole and 
%An unstable orbit for $b=b_c$, in which the photons eternally travel round the black hole. Notice that when $b\le b_c$ the photons can never reach the observer. One can see that this critical value is therefore related to the black hole shadow.

Next, we transform to the cosmological coordinate for
computing a shadow of the black hole seen from an observer.
% extending to the multi black hole cases.
The transformation between the static and cosmological coordinates is given by 
 \cite{kt} as 
$ar=R-M,\, t=T+h(R)$ and 
$dh(R)/dR=-HR^2/[(R-M)V(R)]$. So, from Eq.(\ref{eq:statichorizons}), 
the event horizons in the cosmological coordinate are given by
\begin{eqnarray}
ar_{\pm}=\frac{1}{2|H|}(1\pm\sqrt{1-4M|H|})-M,
\end{eqnarray}
where $r_{+}$ and $r_{-}$ correspond to the cosmological and 
the black hole horizons, respectively. 
%Here, $r_{\pm}$ products of $a$ are constants in $\tau$. 

%Let us consider an observer far from the black hole 
%where the geometry can be asymptotically treated 
%as the de Sitter space time.  
%It should be noticed that the observer can see the shadow only when the 
%universe is expanding, since the cosmological horizon becomes a past horizon 
%and accordingly photons can travel to the observer.  
%On the contrary, if the universe  is contracting, the cosmological horizon 
%is a future horizon and photons are trapped inside the horizon and 
%cannot reach the observer.   

The shape of the shadow is different according to whether the observer
is in the expanding or contracting coordinate.  
In the expanding
coordinate, the observer who is in the asymptotic de Sitter space-time
can see the shadow, since the photons can travel from inside the
cosmological horizon to infinite distance.  While, in the contracting
coordinate, the observer can never see the shadow in the
asymptotically de Sitter space-time but one can see only inside the
cosmological horizon.

Since the colliding black holes must be considered in the
 contracting coordinate,
%It is necessary that the observer see the shadow in a place where the gravity of the black hole does not affect. 
let us consider the situation where an observer is near inside the cosmological
horizon ($r_{obs}\to r_{+}$) in the contracting coordinate. We define
the following parameters, which form the celestial coordinate system, as 
\begin{eqnarray}
\alpha\equiv -\frac{ar_{obs}P^{(\phi)}}{P^{(\tau)}},\quad 
\beta\equiv \frac{ar_{obs}P^{(\theta)}}{P^{(\tau)}},
 \label{eq:ab}
\end{eqnarray}
where $P^{(\mu)}$ are the momenta in the local inertial frame and 
$ar_{obs}$ is the physical distance between the observer and the center of the coordinate. 
%, that is,
Because the shadow's shape of a single nonrotating black hole is a circle 
in the $\alpha$-$\beta$ plane due to the rotational symmetry, we only compute when $\beta=0$.
% setting $P^{\theta}=0$.
Using the transformation from the static to the cosmological coordinate and the critical value of the impact parameter (\ref{eq:impact}), 
%the momenta in the local inertial frame $P^{(\mu)}$ 
%can be rewritten in terms of the momenta in the static coordinate,
%\begin{eqnarray}
%&&P^{(\phi)}=\frac{P_{\phi}}{a\Omega r\sin\theta},
%\nonumber\\
%&&P^{(\tau)}=-a^{-1}\Omega
%\left(\frac{dT}{d\tau}P_{T}+\frac{dR}{d\tau}P_{R}\right).
%\end{eqnarray}
%From Eq.(\ref{eq:enecon}), $P^{R}\to -\sqrt{1+H^2b^2}$ for $R\to\infty$, 
%here the sign of $P^{R}$ is determined by $dR/dT <0$ which is the initial condition of the orbits toward the center $R=0$.
%and taking the limit $r\to \infty$, 
we obtain the critical value of $\alpha$ as 
% a radius of a black hole shadow,
\begin{eqnarray}
\alpha_c= \frac{4M\epsilon}{\sqrt{1+4M|H|}},\quad \epsilon \equiv a|H|(r_{+}-r_{obs}).
\end{eqnarray}
Note that $\epsilon \ll 1$ since we locate an observer near the cosmological horizon.

%where, $\epsilon\to 0$ when taking the limit $r_{obs}\to r_{+}$.
Therefore the shape of the shadow is a circle with this
 radius of $4M\epsilon/\sqrt{1+4M|H|}$ in $\alpha$-$\beta$ space.
One can see that this radius becomes smaller when the observer approaches the cosmological horizon, $r_{obs}\to r_{+}$, due to the geometry on this space-time. 
%%%We set the total mass $M$ to 1 for later convenience.??? it is necessary???

It is necessary to take into account a black hole that is not in the center of the coordinate as more general case. 
Then the black hole moves toward the center of the coordinate in the contracting coordinate. 
We find that its shape is the same as a centered black hole.
%It is merely a transformation of space coordinate although, 
%If the black holes are far enough from each other, the photons come from observer can be already near the de Sitter horizon $1/|H|$
%before the black holes enter the de Sitter horizon. 
%After the black hole cross this de Sitter horizon, only photons which are near the de Sitter horizon and cross the photon sphere of the black hole fall into the black hole. 
%We find that the shape of the shadow of "moving black hole" is disturbed in the movement direction in comparison with "static black hole".
%It is implied that this effect is due to the de Sitter space.

\section{Shadows of colliding black holes}

Let us consider a two black hole system as an example of colliding 
black holes.  It is convenient to adopt the cylindrical 
coordinate ($r,z,\phi$), because the system has
the axial symmetry in this case.  Then the locations of the black
holes are given by $(x_i,y_i,z_i)=(0,0,d_i)$, where $i=1,2$ and we set
$d_1=-d_2$. 
We set an observer at a fixed point inside a cosmological horizon in
the physical coordinate. 
%which will appear after the two black holesmerge. 
%Namely, the distance from the observer to the black holes
%becomes increasing in time in the comoving coordinate.  
For simplicity, we take $\phi_{obs} =0$ and $\theta_{obs}=\pi/2$ 
in terms of the polar coordinate.

Now let us consider ray tracing.   We have to shoot
photons from all the directions to the observer in the contracting 
coordinate to see the shadows of colliding black holes.  
Instead, %due to the technical reason, 
technically it is easier to consider the time reversal of this system.  
Namely, we shoot photons from the observer to all directions in the 
expanding coordinate.   

% We consider photon paths which travel from observer in the expanding
% coordinate instead of paths which travel from outside of the
% cosmological horizon to the observer, since taking time inverse is
% corresponds to changing the sign of inverted Hubble radius $H\to-H$.

%We consider photon paths which travel from observer in the contracting coordinate instead of paths which travel from inside the cosmological coordinate to the observer, 
%since taking time inverse is corresponds to changing the sign of in
%versed Hubble radius $H\to-H$.
Using the parameters $\alpha$ and $\beta$ that have been defined by
Eqs.(\ref{eq:ab}), the initial momenta for photons at the observer
are given by
\begin{eqnarray}
&&P^{r}=-\frac{P_{\tau}}{a^2}\sqrt{1-(\alpha^2+\beta^2)/(ar_{obs})^2},
\nonumber\\
&&P^{z}=\frac{P_{\tau}\beta}{a^3r_{obs}},\quad P^{\phi}=\frac{P_{\tau}\alpha}{a^3r_{obs}^2},\quad {\rm for }\,\,\theta_{obs}=\frac{\pi}{2}.
\end{eqnarray}

The above equations show that the shadows must lie inside the circle 
in the celestial coordinates $\alpha$ and $\beta$ 
with a radius of $ar_{obs}$ ($ \lesssim  ar_{+}$) due to the condition
$1-(\alpha^2+\beta^2)/(ar_{obs})^2\ge 0$.  
We can easily extend the above initial conditions
for arbitrary observers at $\theta_{obs}\ne \pi/2$ by rotating the
two-dimensional vector $(P^{r},P^{z})$ in the $z$-$y$ plane.

We then numerically calculate the photon's geodesic equations from the
observer in the expanding coordinate.  The photons that eventually fall
into the black hole horizons are regarded as shadows.

\begin{figure*}[t]
\begin{center}
%\hspace{-3cm}
 \includegraphics[width=9.5cm,angle=-90]{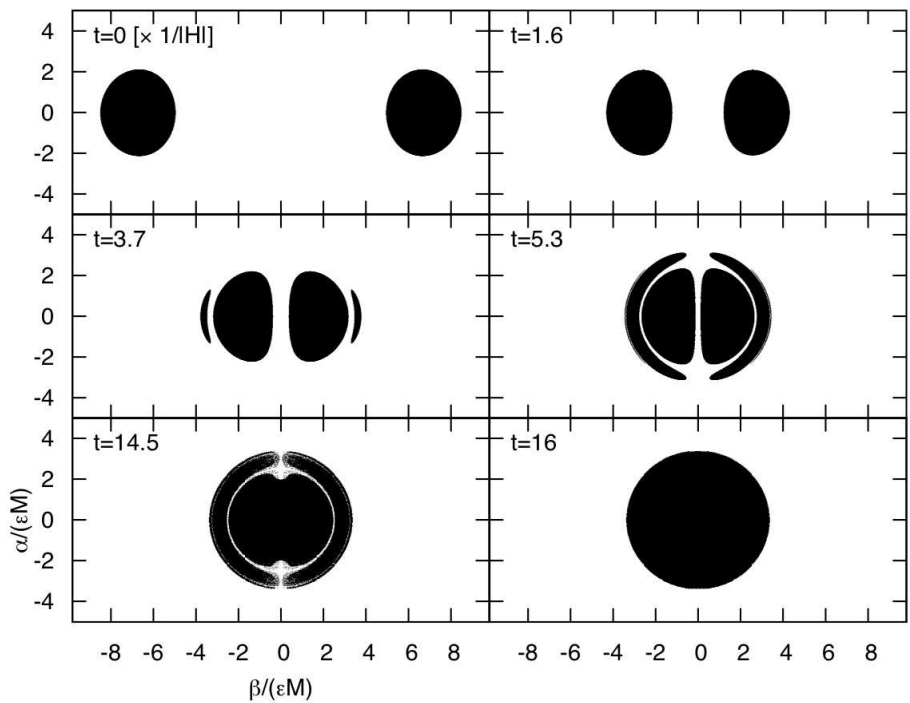}
\end{center}
\caption{Black hole shadows for the two black
hole system plotted in $\alpha$-$\beta$ space normalized by $\epsilon M$
with each physical time $t/|H|^{-1}=0,~1.6,~3.7,~5.3,~14.5,~16$.  Here we have
used the following parameters, $\theta=\pi/2$, $m_1=m_2=M/2$, $H=-0.1/M$ and
$\epsilon=0.01$.}
\label{fig:shadow1}
\end{figure*}

Figure\ref{fig:shadow1} shows that the shadows of two black holes with
same masses $m_1=m_2$ at each physical time $t$ seen by observers at
$z_{obs}=0 (\theta_{obs}=\pi/2)$ and $\phi_{obs}=0$ with $\epsilon=0.01$.  
We take $M=m_1+m_2=0.1/|H|$.
The initial positions of two black holes are $d_1=-d_2=4.5\times 10^{-8}/|H|$. 
Here, the celestial coordinates $\alpha$
and $\beta$ are normalized by $\epsilon M$ in order to keep the shape of the 
shadows independent of a location of the observer.  
  
At $t=0$ and $t=1.6/| H |$, the black holes are still far away from each other.
% so that they are considered as single black holes. 
%As we have previously mentioned, 
However, one can find that their shapes are a little bit elongated in 
the $\alpha$ direction and squeezed in the $\beta$ direction 
from the circles with a radius of 
$4m_i\epsilon/\sqrt{1+4m_i|H|}\sim
1.82\epsilon M$ when they are considered as single black holes in
$\alpha$-$\beta$ space.  This deformation is caused by the existence of 
the other black hole in the opposite side.  

%due to the effect caused by de Sitter space-time. 
%The reason is that, 
%if the black holes are far enough from each other, 
%the photons come from observer can be near the de Sitter horizon $1/|H|$
%before the black holes enter the de Sitter horizon. 
%After the black hole cross this de Sitter horizon, only photons 
%which cross the photon sphere of the black hole 
%fall into the black hole. Therefore, the shapes look like adhering to the de sitter horizon.
%It is implied that this effect is due to the de Sitter space.
%In the figure, One can see that the shadows are near the circumference of the circle with a radius of $1/|H|=10$ in $\alpha$-$\beta$ plane and disturbed in the directions of the movement.
%These are an effect caused by the de Sitter space-time as we have previously mentioned.

At $t=3.7/|H|$, an eyebrowlike structure around each black hole appears.  
% one can find that there are eyelash-like structures around each black hole. 
This kind of structure is quite unique to the multiblack hole system.
The reason why these structures appear is the following.  Let us
consider the winding orbit of a photon around the black hole \cite{luminet}.  
%It is known that this orbit is unstable.  
These orbits form the photon sphere as we have mentioned in Sec. III.  
If the impact parameter of the photon %has slightly more energy 
is slightly smaller than the radius of the photon sphere, 
this photon will eventually
fall into a black hole horizon.  On the other hand, %if the photon has
for a slightly larger impact parameter, % is  little bit less energy, this 
the winding photon will gradually increase the
distance to the black hole and eventually go away from the black hole,
or fall into the horizon of the other black hole.  The latter case
creates the eyebrowlike shadow along the main shadow. The situation is 
similar to the particle motion in the Majumdar-Papapetrou solution \cite{cornish}

% Considering
% a photon in which orbit is unstable, if this photon is deprived of a
% few energy, then the photon will fall into a black hole horizon.  On
% the other hand, if this photon gains a few energy, then the photon
% will go away from the black hole after going round the black hole some
% times.  In this case, the photon can fall into the black hole on the
% other side according to its momentum, therefore it looks another
% shadow.  Satisfying the condition that a photon falls into the
% opposite black hole is limited, the "sub-shadows" only appear around
% "main-shadow". Hence their shapes look like eyelashes.
%It is clear that these sub-shadows are enlarged when the distance between two black holes becomes short.  
%however, if the distance is near enough, two black hole horizons merge before the photon falls, 
%in that case, whether the photon falls into the black hole on this side or the other side does not distinguish, hence sub-shadows and main-shadow look merged 
%as seen by the shadows at $t=16/|H|$ in Fig.\ref{fig:shadow1}.
%In addition, sub-shadows are also enlarged when mass of the black hole on the other side is more massive.
%The black holes finally unit and they looks single black holes having mass $m_1+m_2$. 

At $t=5.3/|H|$, the eyebrowlike structures grow and the main shadows
come close each other.  One can find there still remains a region
that photons can go through between the main shadows.   
The reason why such a region remains is the following.  
In a single black hole system, a black hole horizon is enclosed with the 
photon sphere.  
% Therefore the black hole looks large compared with its black hole
% horizon.  
On the other hand, in a two black hole system, two photon spheres
intersect at the $x$-$z$ plane where the photons cannot fall into
either one of black holes.  Accordingly photons can go through around
this plane, which corresponds to $\beta=0$ in the celestial coordinate 
until two black holes merge and form a horizon. 
Actually, this interaction between two photon spheres cause the deformation of 
black hole shadows at $t=0$, $t=1.6/|H|$, and
$t=3.7/|H|$.

% the photons
% can pass between two black hole horizons if two black hole horizons
% do not merge when the photons r each the center of the coordinate.
% Because the gravitational forces on the photons on a $x$-$z$ plane
% are zero, the shadows at near $\beta=0$ cannot exist until two black
% hole horizons merge.

% Such an exclusive region can be also seen at $t=0$, $t=1.6/|H|$ and
% $t=3.7/|H|$.

At $t=14.5/|H|$, two main shadows have merged and there no longer
exists a region at $\beta=0$ where photons can go through.
This implies that the merger process of two black hole horizons took place 
before the photons reach the center of the coordinate.  
Eventually a shadow of a single black hole appears at $t=16/|H|$.  
The shape of the shadow is a circle with a radius of 
$3.38\epsilon M$ in the celestial coordinate, which corresponds to the 
photon sphere of a single black hole with $M$ as described by Eq. (8).  
%as we have expected.  %Now, there exists a single black hole.
The duration of these merger processes may be estimated as  
$\simeq 15/|H| \simeq 20{\rm hr} (M/10^8M_{\solar})$.

Finally, let us mention a situation when one observes from
arbitrary directions.  We have calculated shadows for several
different values of angle $\theta_{obs}$ at $t=3.7$.  As we decrease
$\theta_{obs}$ from $\pi/2$, the left main shadow of Fig. 1 becomes
elongate, and eventually merges with the eyebrowlike structure of the
right side and forms a ring structure surrounding the right main
shadow \cite{nitta}.

% We calculate shadows with some values of
% $\theta_{obs}$ at $t=3.7$.  Notice that if $\theta_{obs}=\pi/2$, it
% corresponds to the shadows in Fig.\ref{f ig:shadow1} at $t=3.7$.  we
% find that the left large shadow and the right small shadow with
% $\theta_{obs} =\pi/2$ merge, and the small shadow of the left side is
% became smaller as the angle decreases to 0.  Finally, if
% $\theta_{obs}=0$, the shape of the shadow become a circle and a ring.
 
%%%%%%%%%%%%%%%%%%%%%%%%%%%%%%%%%%%%%%%%%

\section{Summary and conclusions}
In this paper, we have calculated photon paths during the collision two black holes and drawn their shadows seen by a distant
observer.  In a realistic case, the merger process is dynamical and
has to be solved by utilizing numerical relativity.  Although this
must be an ultimate goal, instead, we employ the KT solution, which
is the exact solution of the multiblack hole system in the contracting or
expanding coordinate, as a first step.  While we admit the KT solution
is far from the reality, 
[the charge is $Q=10^{46}$esu$(M/10^8M_{\solar})$],
this exact solution enables us to handle
evolution of black hole horizons.  Moreover, it is rather easy %and rigorous 
to calculate the photon paths in this space-time accurately.

% which is an asymptotic space-time of KT solution in far from the black holes. 
% The KT solution can describe black hole merger because they move with
% expansion or contraction.  Of course, calculation of shadows of a
% realistic black hole collision needs a metric which should be
% calculated by numerical relativity. It is too hard to calculate the
% photon path because of necessity of high resolution metric.  The main
% difference between KT-solution and realistic black hole collision is
% whether they have a charge or not, since an observable black hole does
% not have any charge.
%The black holes in KT solution are not affected by any other black holes due to its same charge as mass.
% The realistic collision is mainly induced by their gravity.

% However, it is important because the exact solution enable the
% accurate discussion about photon paths, horizon evolutions without any
% other assumptions, therefore it is useful for extracting the general
% features of black hole collision.  At this point, this solution is one
% of the better model although this solution is not realistic.

We expect that the following two features of black hole shadows obtained
here are general and appear in the realistic situation. The first is
the eyebrowlike structure that shows up during the merger process.
The second is the region on the plane perpendicular to the merger
direction that photons can go through until the last epoch of the
merger.  These features in the shadows can be used as probes to find
the multiblack hole system at the final stage of its merger process.

% exclusive region appear before the two black hole horizons unit.
% These features should be especially emphasized because such a features
% do not exist in the single black hole according to section II.
% Observing these structures of shadows suggests that two black holes
% will collide in the future
%even if we cannot observe the other black hole. 

% Observing a black hole shadow is challenging today.  In particular,
% Sgr ${\rm A}^*$ is a well-known supermassive black hole candidate in
% the Galactic center \cite{ghez,eckart}, and which is a target of the
% black hole shadow observation \cite{falcke}.  One of the most likely
% model of the formation of the supermassive black hole is what increase
% the black hole mass by slow accretion of some black holes.  We expect
% that investigating detailed structures of shadows will suggest some
% informations of the formation of massive black holes in the future
% observation.

%%%%%%%%%%%%%%%%%%%%%%%%%%%%%%%%%%%%%%%%%%%%%%%%%%%%%%%%%%%%%%%%%%%%%%
\section*{Acknowledgments}
%%%%%%%%%%%%%%%%%%%%%%%%%%%%%%%%%%%%%%%%%%%%%%%%%%%%%%%%%%%%%%%%%%%%%%

D.N. would like to thank Shuichiro Yokoyama for useful
discussions. This work was supported in part by a Grant-in-Aid for
Scientific Research from JSPS [No.\,20540280(TC) and
No.\,22340056(NS)] and in part by Nihon University (T.C.). This work
was also supported in part by the Grant-in-Aid for Scientific Research
on Priority Areas No. 467 ``Probing the Dark Energy through an
Extremely Wide and Deep Survey with Subaru Telescope.''
This research has also been supported in part by World Premier
International Research Center Initiative, MEXT, Japan.

%%%%%%%%%%%%%%%%%%%%%%%%%%%%%%%%%%%%%%%%%%%%%%

%%%%%%%%%%%%%%%%%%%%%%%%%%%%%%%%%%%%%%%%%%%%%%%%%%%

%%%%%%%%% references %%%%%%%%%%%%%%%%%%%%%%%%%%%%%%


\begin{thebibliography}{123}

\bibitem{narayan}
J.~Kormendy and D.~Richstone,
  %``Inward bound: The Search for supermassive black holes in galactic nuclei,''
  Ann.\ Rev.\ Astron.\ Astrophys.\  {\bf 33}, 581 (1995); 
  %%CITATION = ARAAA,33,581;%%
R.~Narayan,
  %``Black Holes in Astrophysics,''
  New J.\ Phys.\  {\bf 7}, 199 (2005)
  [arXiv:gr-qc/0506078].
  %%CITATION = NJOPF,7,199;%%

\bibitem{ghez}
A.Ghez, M. Morris, E. E. Becklin, T. Kremenek, A. Tanner,
  %``The Accelerations of Stars Orbiting the Milky Way's Central Black Hole''
  Nature {\bf 407}, 349 (2000)
  [arXiv:astro-ph/0009339].

\bibitem{gillessen} 
S.Gillessen, F. Eisenhauer, S. Trippe,
T. Alexander, R. Genzel, F. martins, T. Ott, Astrophys.\ J.\ {\bf
692}, 1075 (2009)   [arXiv:0810.4674].

\bibitem{kormendy}
J. Kormendy, and  D. Richstone, 1995, Annual Review of 
Astronomy and Astrophysics, {\bf 33}, 581

\bibitem{magorrian}
J. Magorrian, et al., 1998, AJ, {\bf 115}, 2285

\bibitem{merritt}
D. Merritt, L. Ferrarese, 2001, MNRAS, {\bf 320}, L30


\bibitem{sudou}
H.~Sudou, S.~Iguchi, Y.~Murata and Y.~Taniguchi,
  %``Orbital Motion in the Radio Galaxy 3C 66B: Evidence for a Supermassive
  %Black Hole Binary,''
  Science {\bf 300}, 1263 (2003)
  [arXiv:astro-ph/0306103].
  %%CITATION = SCIEA,300,1263;%%




\bibitem{iguchi}
S.~Iguchi, T.~Okuda and H.~Sudou,
  %``A Very Close Binary Black Hole in a Giant Elliptical Galaxy 3C 66B and its
  %Black Hole Merger,''
  Astrophys.\ J.\  {\bf 724}, L166 (2010)
  [arXiv:1011.2647 [astro-ph.GA]].
  %%CITATION = ASJOA,724,L166;%%


\bibitem{falcke}
H. Falcke, F. Melia and E. Agol
  %``Viewing the Shadow of the Black Hole at the Galactic Center''
  Astrophys.\ J.\  {\bf 528}, L13 (2000)
  [arXiv:astro-ph/9912263].

\bibitem{kt}
D.~Kastor and J.~H.~Traschen,
  %``Cosmological multi - black hole solutions,''
  Phys.\ Rev.\  D {\bf 47}, 5370 (1993)
  [arXiv:hep-th/9212035].
  %%CITATION = PHRVA,D47,5370;%%
  
\bibitem{takahasi}
R. Takahashi, Publ. Astron. Soc. Japan, {\bf 57}, 273 (2005). 

\bibitem{gibbons}
G. W. Gibbons and K. Maeda,
% ``Black Holes in an Expanding Universe''
  Phys. Rev. Lett. 104, 131101 (2010)

\bibitem{mp}
S. D. Majumdar, Phys.\ Rev.\ {\bf 72},390 (1947), 
A. Papapetrou, Proc. Roy. Irish. Acad. {\bf 51}, 191 (1947)

\bibitem{nakao}
T.~Chiba and K.~i.~Maeda,
  %``Cosmic hoop conjecture?,''
  Phys.\ Rev.\  D {\bf 50}, 4903 (1994); 
  %%CITATION = PHRVA,D50,4903;%%
K. Nakao, T. Shiromizu and S. A. Hayward,
  Phys.\ Rev.\  D {\bf 52}, 796 (1995)


\bibitem{luminet}
J.-P. Luminet, Astron. Astrophys.  {\bf 75}, 228 (1979). 

\bibitem{cornish}
 C.~P.~Dettmann, N.~E.~Frankel, N.~J.~Cornish,
  %``Fractal basins and chaotic trajectories in multi - black hole space-times,''
  Phys.\ Rev.\  {\bf D50}, R618 (1994).
  [gr-qc/9402027].

\bibitem{nitta}
D. Nitta et al., in preparation.

%\bibitem{ghez}
% A.Ghez, M. Morris, E. E. Becklin, T. Kremenek, A. Tanner,
%   %``The Accelerations of Stars Orbiting the Milky Way's Central Black Hole''
%   Nature {\bf 407}, 349 (2000)
%   [arXiv:astro-ph/0009339].

% \bibitem{eckart}
% A. Eckart, R. Genzel, T. Ott, R. Schoedel,
%   %``Stellar Orbits Near Sagittarius A*''
%   MNRAS, 331, 917 (2002)
%   [arXiv:astro-ph/0201031].

\end{thebibliography}
\end{document}